\documentclass{emulateapj}

\newcommand{\be}{\begin{eqnarray}}
\newcommand{\ee}{\end{eqnarray}}
\newcommand{\beq}{\begin{equation}}
\newcommand{\eeq}{\end{equation}}
\def\simless{\mathbin{\lower 3pt\hbox
      {$\rlap{\raise 5pt\hbox{$\char'074$}}\mathchar"7218$}}}
\def\simgreat{\mathbin{\lower 3pt\hbox
      {$\rlap{\raise 5pt\hbox{$\char'076$}}\mathchar"7218$}}} %> or of order

\begin{document}

\title{Response to ``Concerning Thermal Tides on Hot Jupiters" (Goodman 2009; astro-ph 0901.3279)}

\author{Phil Arras\altaffilmark{1} and Aristotle Socrates\altaffilmark{2} }

\altaffiltext{1}{Department of Astronomy, University of Virginia,
P.O. Box 400325, Charlottesville, VA 22904-4325}

\altaffiltext{2}{Institute for Advanced Study,
Einstein Drive, Princeton, NJ 08540}

\email{ socrates@ias.edu, arras@virginia.edu }


\begin{abstract}
Motivated by the comments of \citet{2009arXiv0901.3279G} on our paper
concerning thermal tides \citep{2009arXiv0901.0735A}, we have studied
an idealized problem to understand the global response of a completely
fluid gas giant planet to thermal forcing at the surface \citep{as2}. Our
findings disagree with the main claims in \citet{2009arXiv0901.3279G}. We
find that significant quadrupole moments can indeed be induced as a
result of thermal forcing. Furthermore, we find that it is possible for
the orientation of the quadrupoles to be such that the planet is torqued
away from synchronous rotation. Given these results, we believe our
proposed thermal tide mechanism \citep{2009arXiv0901.0735A} provides
a viable scenario for generating steady-state asychronous rotation,
inflated radii and possibly eccentric orbits of the hot Jupiters.
\end{abstract}

% -----------------------------------------------------------

\section{The Issue }\label{s: discuss}

Gold \& Soter (1969; GS from here on) originally developed the idea
of thermal tide torques to explain Venus' asynchronous spin rate.
Drawing on their work, \citet{2009arXiv0901.0735A} assesed the
importance of thermal tides for the hot Jupiters. Using the simple
GS prescription for the quadrupole moment, they found that thermal
tides could induce large asynchronous spin, and generate tidal
heating rates more than sufficient to power the observed radii.
\citet{2009arXiv0901.3279G} correctly pointed out that the GS ansatz
does not faithfully represent the fluid motion induced by time-dependent
heating in a completely fluid atmosphere. He argued that the induced
quadrupole moment would be many orders of magnitude smaller than the GS value,
and with an orientation that would act to synchronize the spin, opposite
the GS result.

%Concurrently,
%\citet{2009MNRAS.395..422G} published work reiterating the claim that
%no net quadrupoles can be induced in fluid planets.

Motivated by the criticism of \citet{2009arXiv0901.3279G}, we attempted
to carefully analyze a simplified problem which captures the basic physics
of thermal tide excitation in fluid planets. Our results are presented
in \citep{as2}. From here on,
and unless stated otherwise, all references to equations, figures and
paper sections are to \citet{as2}.

In this note we compare our solutions to the fluid equations (Arras \& Socrates
2009b) to the arguments presented in \citet{2009arXiv0901.3279G}.
As \citet{2009arXiv0901.3279G} is unpublished, we quote the text
from his paper posted on the Cornell University astro-ph archive
(http://arxiv.org/).  We then comment on the accuracy of the GS formula
for the quadrupole moment. Contemporaneous with the posting of
\citet{2009arXiv0901.0735A}, \citet{2009MNRAS.395..422G} published results
on a related problem concerning thermal forcing of hot Jupiters. We
briefly comment on the assumptions and results in these two papers.

\subsection{ \citet{2009arXiv0901.3279G} }\label{ss: Goodman}

The heart of Goodman's argument is contained in the fourth paragraph
on his page 1: ``A jovian planet, being gaseous, lacks elastic
strength. The excess column density of the colder parts of the
atmosphere is counterbalanced‚ to the degree that
hydrostatic equilibrium holds‚ by an indentation of the
convective boundary and a redistribution of the cores mass
toward the hotter longitudes. Insofar as the radial range over which
mass redistribution occurs is small compared to the planetary radius,
the thermal tide therefore bears no net mass quadrupole. The torque on
the atmosphere is opposed by a torque on the upper parts of the
convection zone."

Through these intuitive arguments, Goodman realized that the Gold and
Soter approximation ignores the following fact: though there is a flow from
hot to cold at high altitudes, there is also a return flow at lower
altitudes. In \S 5.1, we explicitly calculate the
pattern of such a flow, by directly solving the fluid equations in the
limit that inertia is ignored. The fact that this basic flow pattern
is not contained in the derivation of the GS formula is a serious
conceptual shortcoming.

The analysis in \S 5.1 examines the limit of zero
forcing frequency, and departs from Goodman's aforementioned paragraph
in two respects. First, in the limit of zero forcing frequency, figure
3 shows that the return flow need not extend as deep as
the convection zone.  The bulk of the fluid motion is confined in the
radiative zone, near the photosphere of the starlight. Second, in \S 5.1
we do not find that density perturbations at high
altitude are compensated by density perturbations of the opposite sign
at lower altitude. Rather, we find that the density perturbation, and
hence torque, is identically zero when fluid inertia is ignored.

Given that density perturbations are zero in the limit of
zero frequency, our next step was to derive finite frequency
corrections. Eq. 45 in \S 5.2 shows that
finite forcing frequency corrections give rise to a nonzero quadrupole
moment. Have we violated hydrostatic balance, assumed by Goodman, by
including finite frequency?  The answer rests on a technical detail, which
may be important in future investigations. The equation of hydrostatic
balance $dP/dz=-\rho g$ is obtained from the radial momentum equation
by throwing away the inertial terms. Even if we were to throw away
the inertial terms in the radial momentum equation, but kept them
in the horizontal momentum equations, we would still find a nonzero
quadrupole moment of the correct sign, although its magnitude would
be slightly different (throwing away the $-1$ in the parenthesis in
eq. 45 changes the prefactor from $4$ to $5$).

By what factor should the GS quadrupole moment be reduced
due to the ``isostatic compensation" from the return flow?
\citet{2009arXiv0901.3279G} argued above  that the reduction factor
should be a power of $H/R$. By contrast, solution of the fluid equations
in the limit of small forcing frequency, ignoring gravity waves,
finds a frequency dependent reduction factor $\sim 4(\sigma/N)^2$ (see
eq. 40).  Allowing for gravity waves, the calculations in
figures 4 and 5 show that the response
is larger than eq. 45 by 1-3 orders of magnitude in
the relevant period range 1 day - 1 month, due to the excitation of
gravity waves.

\citet{2009arXiv0901.3279G} clarifies the range of forcing period over
which the quadrupole moment should be isostatically compensated on his
page 2: ``Whereas terrestrial isostasy operates on such long timescales
that rock behaves as fluid, the corresponding timescale for gaseous
planets is dynamical, hence less than the tidal period."

One of the key results of Arras \& Socrates (2009b) is that low 
radial order gravity
waves dominate the overlap with the thermal tide forcing. Hence the low
frequency limit in eq. 45 does not apply until forcing
periods $\sim 1$ month, comparable to or longer than the forcing periods
of interest (Arras \& Socrates 2009a). Note that this surprising
result is completely different from the case of gravitational forcing
of incompressible fluid bodies, where the low frequency limit applies
below the characteristic dynamical frequency $(GM_p/R_p^3)^{1/2}
\sim {\rm (hours)^{-1}} $ for a gas giant planet.

Lastly, \citet{2009arXiv0901.3279G} discusses the orientation 
of the induced quadrupole on his page 2:
``Thus, the tidal torque
claimed by Arras \& Socrates (2009) vanishes to first order in the
density variations of the thermal tide. To the next order, the
quadrupole moment of the thermal tide aligns with the hottest and most
distended parts of the atmosphere, because mass elements are weighted
by the squares of their distances from the center.  This will lead to
a torque of the opposite sign to that of $\Delta \Omega$, hence
driving the planet toward synchronous rotation. Similarly, the phase
lag of the thermal tide associated with an orbital eccentricity will
affect the orbit only to second order, and will tend to circularize
the orbit."

The analytic solution to the fluid equations in the low frequency
non-resonant limit (eq. 45) has the correct sign to
drive asynchronous rotation, contrary to Goodman's claim. Including
the effect of gravity waves, figures 4 and
5 show that the sign of the quadrupole can alternate with forcing
frequency. These sign changes are due to both the
Lorentzian factors in eq. 60, as well as the signs of the
quadrupole moments for individual modes (figure 6). 
Even in this more complicated case, frequency
ranges still exist where the thermal and gravitational tide torques
may oppose each other, leading to an equilibrium spin state.

In summary, \citet{2009arXiv0901.3279G} correctly points out the
deficiencies in the Gold and Soter approximation employed by Arras \&
Socrates (2009a). However, the solutions to the fluid equations
presented in \citet{as2} differ both qualitatively and quantitatively from the
basic picture outlined in his work.   
Consequently, we disagree with Goodman's
criticism of Arras \& Socrates (2009a) i.e., that thermal tides cannot
lead to asynchronous spin and eccentric orbits.

\subsection{the Gold \& Soter approximation, and the calculations
of Arras \& Socrates (2009a) }

Gold and Soter's ansatz involves a major assumption: that the fluid
elements remain at roughly constant pressure, so that density
perturbations are related to temperature perturbations by $\delta
\rho/\rho = - \delta T/T = - \Delta s/c_p$. From eq. 29,
we see this is indeed true if the $\delta p$ and $\xi_r$ terms can be
ignored. For low frequency forcing, ignoring the $\delta p$ term may
be a good approximation, but for fluid atmospheres we have seen it is
{\it not} a good approximation to ignore the $\xi_r$ term. If there is
a solid surface, and the boundary condition at this surface is
$\xi_r=0$, then the Gold and Soter ansatz may, in fact, hold. This
might be realized if the heat was all deposited at or near the solid
surface, rather than well above the surface.

On a more practical level, for fluid planets with surface radiative layer,
we have found the Gold and Soter approximation overestimates the quadrupole moment
and torque by more than an order of magnitude for the calculations in this paper.
Since the steady state planetary radii powered by tidal heating found by
Arras \& Socrates (2009a) were rather large compared to observed planets,
the reduction in torque found in this paper may bring their theory into better
agreement with observations.

To summarize, we have found that the Gold and Soter ansatz for the
quadrupole moment is qualitatively, but not quantitatively correct. It may be
viewed as a convenient order of magnitude estimate.

\subsection{ \citet{2009MNRAS.395..422G} }

\citet{2009MNRAS.395..422G} studied perturbations to a hot Jupiter
atmosphere induced by time-dependent radiative heating due to asynchronous
rotation.  Their primary result is that waves excited by the thermal
forcing can radiatively damp and transfer angular momentum vertically
in the atmosphere.

\citet{2009MNRAS.395..422G} did not study if net quadrupoles could be
induced in the atmosphere. By working in plane parallel geometry and
requiring the pressure perturbation to vanish at the base of the grid,
they set the quadrupole to zero by hand. As the goal of their paper
was to study differential rotation induced by the damping of downward
propagating gravity waves, it is likely a good approximation to work
in plane parallel geometry, ignoring the possible existence of net
quadrupoles.  Note that Arras \& Socrates (2009b) focus on the
complementary issue of net quadrupoles, ignoring possible driving of
differential rotation.

% -----------------------------------------------------------

\section{Summary}
\label{s: summary}

In summary, the results in this paper confirm that quadrupole moments
of the correct sign and approximately correct magnitude may be induced
by time-dependent insolation, confirming the basic assumptions of
Arras \& Socrates (2009a). A future study will use the results
from this paper in concert with a thermal evolution code for the hot
Jupiters (Arras \& Bildsten 2006; Arras \& Socrates 2009a) to construct
more detailed steady state solutions for the planetary rotation and radius, and
the orbital eccentricity.

%--------------------------------------------------------------------------------

\acknowledgements We thank Peter Goldreich for helpful discussions.
Also, we thank Jeremy Goodman, Gordan Ogilvie and Pin-Gao Gu for
raising the issue of isostatic adjustment.

%--------------------------------------------------------------------------------

%--------------------------------------------------------------------------------

\end{document}